# Collaborative effects of wavefront shaping and optical clearing agent in optical coherence tomography

Hyeonseung Yu,[a,f] Peter Lee,[b,f] YoungJu Jo,[a,f] KyeoReh Lee,[a,f] Valery V. Tuchin,[c,d,e] Yong Jeong,[b,f] YongKeun Park [a,f,*]

[a]Department of Physics, Korea Advanced Institute of Science and Technology, Daejeon 34141, South Korea
[b]Department of Bio and Brain Engineering, Korea Advanced Institute of Science and Technology, Daejeon 34141, South Korea
[c]Research-Education Institute of Optics and Biophotonics, Saratov National Research State University, 83, Astrakhanskaya Street, Saratov 410012, Russia
[d]Institute of Precision Mechanics and Control of Russian Academy of Sciences, Saratov 410028, Russia
[e]Laboratory of Biophotonics, National Research Tomsk State University, Tomsk 634050, Russia
[f]KAIST Institute of Health Science and Technology, Daejeon 34141, Republic of Korea

**Abstract**. We demonstrate that simultaneous application of optical clearing agents (OCAs) and complex wavefront shaping in optical coherence tomography (OCT) can provide significant enhancement of the penetration depth and imaging quality. OCA reduces optical inhomogeneity of a highly scattering sample, and the wavefront shaping of illumination light controls multiple scattering, resulting in an enhancement of the penetration depth and signal-to-noise ratio. A tissue phantom study shows that concurrent applications of OCA and wavefront shaping successfully operate in OCT imaging. The penetration depth enhancement is further demonstrated for ex vivo mouse ears, revealing hidden structures inaccessible with conventional OCT imaging.

*Address all correspondence to: YongKeun Park**, E-mail: yk.park@kaist.ac.kr

## 1  Introduction

Developing a deep-tissue imaging tool with good optical resolution is one of the important goals of biophotonics because medical diagnosis usually requires imaging of fine details of whole tissue structures. Optical coherence tomography (OCT) is an extensively developed imaging tool for biological tissues. Based on low-coherence interferometry, OCT provides micrometer scale resolution and enables non-invasive high-speed imaging[1]. OCT is widely used in ophthalmology because it can fully access retina, choroid and sclera layers of human eyes *in vivo*.[2, 3] OCT has a higher penetration depth of 1–2 mm compared to other optical imaging modalities such as multi-photon microscopy and confocal microscopy.[4] Photoacoustic microscopy also offers deep tissue imaging capability but optical absorption contrast is used instead of elastic backscattering signals[5]. Despite its advantages, OCT cannot provide penetration depth on centimeter scales, as is possible with ultrasound imaging.

The limitation of the penetration depth of OCT originates from multiple light scattering caused by inhomogeneous refractive index distributions in biological tissues. The strategy of OCT imaging is to deliver light deep into a target region and measure backscattered light. However, light is severely scattered during propagation inside a biological tissue. Biological tissues are composed of various cellular components, each having different chemical compositions and thus different refractive indices. For example, cellular organelles, cytoplasm, and collagen fibers have refractive indices higher than that of interstitial fluid. More quantitatively, optical waves have a reduced scattering coefficient of about one $mm^{-1}$ in most biological tissues.[6] This implies that after propagation of 1 mm, the light experiences a few scattering events (on average ten for the scattering anisotropy factor typical for tissues, $g = 0.9$), after which the light completely loses its original propagation direction. To minimize the effect of multiple scattering events, OCT uses coherence gating to collect scattering signals selectively with the same optical path lengths. However, the amount of single scattered light exponentially decays as the depth increases, and the multiple scattering becomes dominant beyond a few mean free paths.[7] Thus, it is extremely challenging to image scatterers at depths greater than approximately 1 mm; thus, penetration depth is limited to a few millimeters.

One method of visualizing deep inside tissues in OCT systems is to utilize optical clearing agents (OCAs) to compensate for optical inhomogeneity.[8, 9] The primary role of OCAs is to reduce refractive index mismatches among tissue structures and thus to reduce light scattering in the tissues.[10-12] Due to the complexity of tissue structures, the reduction of the scattering cannot be explained merely by the refractive indices of the OCAs[13, 14]. In fact, various mechanisms are involved, such as the dissociation of collagen[15, 16], the ability to disrupt the hydration shell and the water-mediated hydrogen bonds of collagen molecules,[17] and the dehydration process[10, 18-20]. In addition to penetration depth enhancement for tissue imaging[21-23], OCAs also enable optical blood clearing [24] and monitoring of molecular diffusion[25-30] using OCT systems.

Whereas OCA approaches use the scattering properties of tissue samples for the enhancement of the penetration depth, it has been recently demonstrated that controlling the illumination light in optical imaging can offer a high possibility of overcoming multiple scattering without affecting the sample. Although adaptive optics approaches were exploited to correct aberrations, the multiple light scattering requires more degrees of freedom[31, 32]. The formation of optical focus can be interpreted as constructive interference of scattered light waves. Multiple light scattering scrambles phase information of an impinging beam, and thus the desired constructive interference, or optical focus, cannot be achieved deep in a tissue. However, these perturbations can be adjusted by controlling the complex field of the incident light [33, 34]. This idea was first demonstrated by showing that clear optical focus can be formed through highly scattering media[35]; later, this wavefront shaping approach was extended to optical systems that are based on OCT principles. The focusing of light into deep tissue was demonstrated using optical coherence microscopy[36], and penetration depth enhancement was studied in a spectral domain OCT system[37-39], the so-called wavefront shaping OCT (WS-OCT). The measurement of a time-resolved reflection matrix based on the principles of time-domain OCT enabled selective focusing inside tissues[40] and diffraction-limited imaging[41, 42]. In a related field, wavefront shaping approaches have also been exploited for deep-tissue imaging with multi-photon microscopy[43, 44]. One of the advantages of using wavefront shaping in OCT is that the light control can also compensate system aberrations or the degradation in resolutions at out-of-focus planes[45].

In this work, we demonstrate that simultaneous applications of wavefront shaping and optical clearing agents can provide collaborative effects for enhancing penetration depth using WS-OCT. Even though both the wavefront shaping and OCAs enhance the penetration depth, the collaborative effect is not straightforward due to the loss of the scattering contrast by OCAs. Because the wavefront shaping technique utilizes the backscattered signals for the optimization, the reduced scattering by OCAs should be still strong enough so that the wavefront shaping works. A tissue-mimicking phantom and an *ex vivo* mouse ear tissue were visualized with the application of 70% glycerol solution and wavefront shaping via a digital micro-mirror device (DMD). Benefitting from balanced applications of the OCAs and the wavefront shaping, collaborative effects of both methods offer unprecedented enhancement of the penetration depth compared to that possible with conventional spectral domain OCT (SD-OCT) imaging.

## 2 Principles

The principles of OCA and wavefront shaping in OCT are illustrated in Fig. 1. In a conventional OCT system, an uncontrolled beam is incident on a sample and backscattered signals from a scatterer are used as image contrast. The desired situation is one in which a tight focus is formed at a target axial depth and signals are strongly reflected from the focus, providing vivid images of tissue structures. However, due to optical inhomogeneity inside the tissue, the beam propagation paths that are supposed to be focused on the target spot experience spatially varying perturbations.

This situation leads to severe degradation of the optical focus, as illustrated in Fig. 1(a). There are two different approaches that can alleviate this degradation of the focus: the application of OCA, as shown in Fig. 1(b), and wavefront shaping, as shown in Fig. 1(c). In Fig. 1(b), OCA can be seen to reduce the inhomogeneity of the tissue, and the multiple scattering becomes less severe. The penetration depth can be extended at the expense of the tissue composition change. In contrast to the OCA approach, which directly changes the sample, the wavefront shaping method controls light propagation in tissues to maximize the light delivery to scatters and the reflection from the scatters, as shown in Fig. 1(c). Because the wave-propagation in tissues is a deterministic process, we can find an optimal wavefront that can selectively focus at the target depth inside the tissue. While the OCA directly suppresses multiple light scattering by manipulating the tissue, wavefront shaping controls the light paths without affecting the tissue structure. Therefore, these two approaches can work concurrently, and the combination of both methods provides collaborative effects to control multiple light scattering. This best situation of the combined approach is illustrated in Fig. 1(d). For all the above cases, the enhancement of the penetration depth can be maximized when the OCA technique and the wavefront shaping approaches are simultaneously applied.

## 3 Experimental Setup

A schematic of the experimental setup is shown in Fig. 2. The system was adapted from an SD-OCT system by incorporating high-speed wavefront modulation[37]. In a conventional system, a collimated beam from an illumination source directly passes through a beam splitter and the divided beams arrive incident on a sample and a reference mirror, respectively. For the wavefront shaping OCT (WS-OCT), a wavefront shaping unit consisting of a digital micro-mirror device (DMD) and a dispersion compensating grating (G1) is additionally installed, as indicated by the gray dashed rectangular area. After the wavefront shaping unit, OCT signals are acquired and processed based on SD-OCT principles. Detailed theories on SD-OCT systems can be found in the literature[45].

In the WS-OCT system, a broadband superluminescent diode (SLD-52, Superlum Diodes, Ltd., Ireland) with a center wavelength of 1025 nm (full width at half maximum = 100 nm, 9.7 mW) is used as the light source. The beam propagates in free space after the collimator and impinges on the DMD (0.7″ XGA, 1024×768, Texas Instruments, United States). The DMD is a 2-D array of 800K micro-mirrors; each mirror directs the incident beam at 12° or -12°. The light redirected from the DMD with the angle of 12° propagates along the optical axis; the light redirected with the angle of -12° is blocked out. Thus, the control of individual micro-mirrors provides spatially structured incident beams. Due to the pixelated periodic structure, the DMD essentially works as a grating. For broadband illumination, the dispersion is produced and therefore is compensated for by the grating (300 lines/mm, GR25-0310, Thorlabs, United States). The modulated field pattern is delivered to the back pupil of an objective lens (LSM-03BB, Thorlabs, United States) through two 4-$f$ systems. The modulated incident beam is divided into a sample beam and a reference beam at the beam splitter, and then the reflected light from the sample and the mirror are collected by a single mode fiber. At the output facet of the collection fiber, interference signals of the sample and reference beams are analyzed by a spectrometer system consisting of a grating (1200 lines/mm, Wasatch Photonics, United States) and a line CCD (SU1024, 1024 lines, 92 kHz frame rate, Sensors Unlimited Inc,. United States). A 2-D image is acquired by steering the sample beam with one mirror of a 2-axis glavo mirror (6210H, Cambridge Technology, United States) while the other mirror is unused. Because the DMD plane is conjugated with the scanning mirror plane, the amplitude modulation at the DMD plane is converted to full-field modulation at the sample plane. In the setup, the reference beam is also modulated by the DMD in order to preserve the simplicity of the setup. It is possible in principle to separate the reference beam and sample beam, but matching the optical path lengths between the separated sample and reference beams requires more complicated setup. We also considered that simultaneous modulation of both arms does not result in the difference in the acquired signal, because the interference signal is a function of the temporal frequency, whereas the DMD modulates the spatial profile of the incident beam. The spatial modulation of the reference beam only changes the coupling efficiency to the single mode fiber, which is calibrated in the signal processing step.

The goal of wavefront shaping optimization is to maximize the reflected signal from the desired depth. Because tissue compositions are entirely unpredictable, characterization of the optical responses of the tissues is required to find the optimal wavefront. Furthermore, tissue structures vary spatially. Thus, the optimization is sequentially processed for each A-scan. In the optimization of a single A-scan, the detailed procedures are divided into three stages: 1) recording of OCT signals with Hadamard pattern projections to characterize the tissue response, 2) calculation of optimal DMD patterns, and 3) recording of enhanced signals for projections of optimal DMD patterns.

At first, 7,500 2-D Hadamard patterns are sequentially projected on the DMD and A-scan profiles are recorded for each pattern. Hadamard patterns are mutually independent, and this property minimizes the redundancy of the image acquisitions. Second, we select 200 depth positions as regions of interest (ROI) in the A-scan and individually calculate the optimal pattern at each depth. Here, the axial resolution is 2.9 µm and 200 pixels ROI corresponds to a 580 µm depth range. In the calculation of the optimal DMD pattern at a specific depth, we selectively choose 500 Hadamard patterns out of the 7,500 initial patterns that give the largest signal at the given depth. Because the coherent control of light scattering exploits the linear relation between the input and reflected light fields, raw signals are manipulated in a linear scale. After that, we coherently sum the selected 500 patterns to produce a single optimal pattern. Consequently, we have a total of 200 optimal patterns that individually generate the optimal signal at each depth of the ROI. Third, the calculated 200 optimal patterns are projected sequentially on the DMD, and the enhanced OCT signals are recorded. The recorded 200 A-scan signals contain the enhanced signals at different depth positions, so they are compounded into a single A-scan image in which all pixels in the ROI have the enhanced signal. The acquisition time of a single optimized A-scan image is 15 seconds, and 25 A-scans are acquired to obtain a 2-D image.

To demonstrate the coherent control effect of the wavefront shaping, we also acquired the images using the spatial compounding method, which provides a way of suppressing coherent speckle noises by incoherently averaging several OCT signals from spatially- or temporally-varying illuminations[46-49]. In our case, 25 random binary DMD patterns were illuminated on the sample, and the 25 acquired OCT signals were incoherently summed to reduce the noise. To properly compare the spatial compounding method and the wavefront shaping, the size of the macro-pixels in the DMD and the camera acquisition settings are set to be the same for both cases.

## 4  Tissue phantom Study

To validate the idea of the combined effect of OCA and wavefront shaping for the penetration depth enhancement, we first conducted experiments on a tissue phantom fabricated with 20% intralipid and agar[50]. Among various tissue phantom models [51], the used phantom has several advantages: 1) the good permeability of agarose enables the efficient diffusion of OCAs, 2) the scattering coefficient of the tissue phantom is easily tuned by changing the weight/volume fraction of the intralipid

### 4.1  Sample preparation

For the fabrication of the tissue phantom, agarose solution is first prepared by dissolving 1 g of agarose powder (A5304-100G, Sigma-Aldrich, United States) in 100 ml of distilled water at 95ºC. Then, 20% intralipid (Fresenius Kabi AB, Sweden) and agarose solution are homogenously mixed at a 2% weight/volume fraction at 75ºC. After the temperature of the mixture reverts to room temperature, the mixture becomes a solid state spontaneously. The reduced scattering coefficient of the tissue phantom is measured and found to be $1.21\pm0.13$ mm$^{-1}$; this value is obtained by integrating the sphere measurement and inverse-adding doubling methods [52]. Most biological tissues have the reduced scattering coefficients of approximately one mm$^{-1}$, which is comparable to the value of the fabricated tissue phantom [6]. We note that the scattering properties of the use tissue phantom are not identical to those of biological tissues. The anisotropy factor of intralipid is 0.52 at 1064 nm wavelength[53]. In general, biological tissues have anisotropy factors of 0.8—1.0.[6] Glycerol, a 70% solution in water, was used as an OCA. The refractive indices of intralipid and agarose are 1.357[53] and 1.3348[54] at 1064 nm, respectively. From the law of Gladstone and Dale, the mean refractive index of the tissue phantom is estimated as 1.3392. The scattering particles in intralipid are soybean oil, and its refractive index is 1.460[12]. When the glycerol with the RI of 1.463 is applied to the tissue, it is expected to match the refractive indices of the background and soybean oil by increasing the refractive index of the background.

The target imaging sample has a layered structure: layer I is the tissue phantom, layer III is a translucent petri dish and the layer II is the boundary between layer I and layer III, as shown in Fig. 3(a).

*4.2 Experimental results*

To demonstrate the effect of the OCA and the wavefront shaping, we compared wavefront shaping and conventional images of the tissue phantom before applying OCA and after applying OCA. A photograph of the sample before the application of the OCA is shown in Fig. 3(b). The tissue phantom layer is fabricated to have a slope so that the left part is thinner than the right part. Due to the small thickness, the left part looks more grayish than the right part, but there is still a clear difference in the color between the left part of the sample and the petri dish. When the OCA was applied to the left thinnest part, reduction in scattering made the layer almost transparent, and the petri dish below became visible, as shown in Fig. 3(c). This clearing effect is more clearly visible in the insets with the magnified images of the edge. Before applying the glycerol OCA, the wavefront shaping and uncontrolled illumination results are acquired as depicted in Figs. 3(d-g). The imaging region is set to be around the red dashed circle in Fig. 3(b). Fig. 3(d) shows an image with wavefront shaping; Fig. 3(e) shows an image acquired using uncontrolled illumination; Fig. 3(f) is an image acquired using the spatial compounding method. The images of Figs. 3(d-f) were all acquired from the same incident beam power of 0.55 mW. The images obtained using the uncontrolled illumination and spatial compounding methods show similar penetration depths and signal levels, but we can see that these two conventional imaging methods have large shortcomings compared to the wavefront shaping method regarding penetration depth and signal to noise ratio (SNR). For further comparison, the image in Fig. 3(g) was acquired using uncontrolled illumination but with doubled incident power of 1.1 mW. The wavefront shaping image shows a comparable level of SNR and even slightly more features below layer II compared with the sample subject to doubled power illumination. To quantitatively compare the signal levels for the four cases, we plotted the averaged signal profiles over 25 A-scans versus depth. We can see that wavefront shaping yields the largest signal over the depth ranges. The signal levels exponentially decay as the depth increases, so layer I is the only accessible imaging area even in wavefront shaping.

After demonstrating the enhanced penetration depth with wavefront shaping in the original sample state, we next applied the OCA to the same region and acquired images with the same procedures as used above. After adjusting only the z-position of the sample stage, we applied 300 µl of 70% glycerol solution to the imaging region and restored the sample to the original position. We waited for 10 minutes after the application of the OCA to allow OCA to diffuse into the sample. A wavefront shaping image acquired after the OCA is shown in Fig. 3(i). The third layer became visible; it had not been possible to visualize this layer using any of the previous methods. Also, the signal level decreased in the phantom layer, which implies that the OCA successfully penetrated into the tissue phantom and made the sample more transparent. This effect can also be observed in the uncontrolled beam image and the spatial compounding image, as shown in Fig. 3(j) and 3(k), respectively. Since the scattering of the phantom sample is significantly reduced, no meaningful signal having structural information is observed in layer I. Fig. 3(l) shows an image acquired using double powered uncontrolled illumination.

The average profiles over 25 A-scans are plotted in Fig. 3(m). Compared to the signal profiles obtained before the OCA, the signal level has decreased in layer I and the signals from the petri dish in layer III have increased against the background signal level. Without the OCA, the maximum penetration depth for wavefront shaping is less than 300 µm, as can be seen in Fig. 3(h). With the OCA and wavefront shaping, structures deeper than 600 µm from the surface became accessible, as can be seen in Fig. 3(m), which shows a more than two-fold enhancement in the penetration depth.

## 5   Ex vivo mouse ear

To test the suitability of our technique for real biological samples, we next applied the OCA and wavefront shaping to an *ex vivo* mouse ear. In the tissue phantom study, the optical inhomogeneity of the petri dish was not affected by the application of the OCA, and these preserved scattering contrasts were easily imaged using WS-OCT. For real biological samples, however, all imaging regions are optically cleared by the OCA. Therefore, it is important to confirm the proposed technique in real biological tissues.

## 5.1 Sample preparation

After anesthetization with an intraperitoneal injection of Tiletamine-Zolazepam and Xylazine mixture (30:10 mg/kg body weight), a mouse ear was dissected from a specific pathogen-free C57BL/6J mouse (Jackson Laboratory, U.S.A.). All experimental procedures were approved by the Institutional Animal Care and Use Committee (IACUC) of the Korea Advanced Institute of Science and Technology (KAIST). After the dissection, the ear was immersed in 70% glycerol solution for an hour. A photograph of the whole mouse ear is shown in Fig. 4(a); the representative histology of the sub-region is shown in Fig. 4(b). To compare the histology and the OCT images, subsequent OCT signals are plotted according to physical depth while assuming a refractive index of 1.4 for the mouse skin [6]. For example, a 140 μm optical depth in OCT signal is converted to a 100 μm physical depth. This scale factor can vary upon the application of OCAs. Since the refractive index of 70% glycerol is 1.463, the change in the scale factor is less than 5% even under the assumption that the sample medium is completely replaced with glycerol. This change corresponds to the 12 μm for a 300−μm−thickness sample. Thus we assume there is no significant effect of the OCA on the comparison between the histology and the OCT images.

## 5.2 Experimental results and Discussion

Three different locations on the mouse ear were imaged before OCA, and three other locations were imaged after the application of OCA. The OCT images before OCA are first presented in Figs. 4(c-h). For clear comparisons, we show here only the wavefront shaping images and the uncontrolled beam images for the same incident power of 0.55 mW. The wavefront shaping images [Figs. 4(c,e,g)] show superficial layers of the ear tissue, while image signals cannot be clearly seen with the uncontrolled beam illumination, as shown in Figs. 4(d,f,h). Although wavefront shaping provides enhanced penetration depth and SNR over those characteristics of uncontrolled beam illumination, only a single layer is visible, and the structural information of the ear is hardly attainable even in the wavefront shaping case.

To achieve more enhancement of the penetration depth using OCA, we immersed the ear in the 70% glycerol solution and acquired OCT images, which are shown in Figs. 4(i-n). Since the whole ear was taken off of the sample stage and immersed in the OCA solution, locating the original imaging positions was impossible, so we assumed that the penetration depth was constant for nearby imaging areas. As can be clearly seen in Figs. 4(i,k,m), the wavefront shaping technique provides remarkable enhancements: the double layered structures can be visualized for every location. For the uncontrolled input beam cases [Figs. 4(j,l,n)], the signal strengths are slightly better than those acquired before the OCA, but still they provide blurred images of the tissue structures.

Figure 4(b) provides a representative histology image taken from the middle ear of a mouse that may reflect the multilayer structure found in Fig. 4(i,k,m). It should be noted that the histology and OCT images were acquired from different ears. However, we assumed that the general structure is about the same in ears of the same species. In particular, for the case shown in Fig. 4(i), we speculate that the three layers (I, II,III) match the epidermis, auricular cartilage, and smooth muscle, as found in Fig. 4(b).

To compare these data quantitatively, we plotted the averaged depth profiles over 25 A-scans in Figs. 5(a-f) for each location, as can be seen in Fig. 4. The same trends as those found in the cases of the tissue phantom study are observed: 1) wavefront shaping gives the highest signal level in all locations, 2) the uncontrolled beam illumination with 0.55 mW and the spatial compounding method with 0.55 mW give the smallest SNR. When the wavefront shaping cases are considered exclusively, imaging depths are limited up to approximately 200 μm before the OCA and the penetration depth reaches about 300 μm after applying the OCA.

The results above clearly show the collaborative effects of wavefront shaping and OCA in OCT. Using scattering phantom samples and *in-vitro* biological tissue samples, we have demonstrated that both the penetration depth and the signal-to-noise ratio can be significantly improved, compared to the case of using conventional OCT without OCA treatment. Importantly, the simultaneous application of wavefront shaping and OCA results in collaborative effects for deep-tissue optical imaging.

For future studies, the optimal conditions of the OCA treatment and wavefront shaping should be further investigated. For example, various optical clearing agents are available for different tissues types and imaging conditions.[19, 55, 56] The optimal selection of OCAs will provide more enhanced penetration depths. The wavefront shaping technique also has much room for improvement. In our previous results [37], the depth enhancement factor increases as the scattering of the tissue phantom is reduced. This trend shows a positive sign that further optimization of collaborative effect can lead to the synergetic effect of the both method, where the depth enhancement is greater than the sum of individual effects. Although we focused on the reduced scattering coefficient of the sample, the effects of the variation on the anisotropy factor and the scattering coefficient is still an open question. Proper setting of the wavefront parameter according to the sample conditions is expected to bring about higher enhancements of the penetration depth.

In the current approach, there is partial loss of sample information due to OCAs, as seen in the optical clearing of the phantom layer in the tissue phantom study. Furthermore, there should be balanced applications of OCAs. The wavefront shaping utilizes backscattered signals as a feedback signal. Thus the reduction of scattering can prevent efficient wavefront shaping. For example, the complete clearing of OCAs makes the present approach inapplicable. The minimum amount of scattering for collaborative effects should be further studied.

The ultimate goal of deep-tissue imaging is to achieve penetration depth enhancement *in vivo*[23, 57, 58]. We anticipate that *in vivo* application can be realized in the near future because OCA and WS-OCT have worked successfully in *in vivo* environments. However, many experimental difficulties have to be overcome. First, the delivery of the OCA into living tissues is not straightforward and usually requires special treatments.[21, 59] Second, the image acquisition speed with wavefront shaping must be further optimized. Recently, we demonstrated that the wavefront shaping OCT works for static biological tissues *in vivo*[39]. However, the 15-second optimization time for a single A-scan is too slow to image moving tissues. To be applicable for *in vivo* imaging, the wavefront shaping should be fast enough as the aberration correction in sensorless adaptive optics OCT[60, 61]. To enhance the optimization speed, various approaches can be applied, such as the computation time reduction using graphics processing unit (GPU), implementing the wavefront shaping in swept-source OCT platform or designing efficient optimization algorithm based on pre-acquisition plain OCT data. If these problems are adequately addressed, wavefront shaping and OCAs will provide a promising method for *in vivo* deep-tissue imaging.

## 6 Conclusion

Here, we have demonstrate penetration depth enhancement in WS-OCT by simultaneously using wavefront shaping and OCAs. With the application of 70% glycerol solution as an OCA and the wavefront shaping technique via a DMD, we visualized multi-layered structures of a tissue phantom and *ex vivo* mouse ear with unprecedented penetration depth enhancement in comparison to that possible with conventional OCT imaging. For the image optimized with wavefront shaping, the application of OCA provided an additional enhancement factor of 1.5–2 for the penetration depth.

*Acknowledgments*

This work was supported by KAIST, and the National Research Foundation of Korea (2015R1A3A2066550, 2014K1A3A1A09063027, 2012-M3C1A1-048860, 2014M3C1A3052537) and Innopolis foundation (A2015DD126). VVT was supported by Russian Presidential grant NSh-7898.2016.2 and the Russian Government grant 14.Z50.31.0004.

**Hyeonseung Yu** received BS degree in Physics and Mathematical Science from KAIST. He is currently a PhD student at KAIST, in the Department of Physics.

**Peter Lee** received BSE and MSE degrees in Bioengineering from University of Pennsylvania.
He is currently a PhD student at KAIST, in the Department of Bio and Brain Engineering.

**YoungJu Jo** is currently a BS student at KAIST, in the Department of Physics.

**KyeoReh Lee** received a BS degree in Physics from KAIST. He is currently a PhD student at KAIST, in the Department of Physics.

**Valery V. Tuchin** is a professor and chairman of optics and biophotonics at Saratov State University. He is also the laboratory head at the Institute of Precision Mechanics and Control, RAS. His research interests include biophotonics, tissue optics, laser medicine, tissue optical clearing, and nanobiophotonics. He is a member of SPIE, OSA, and IEEE. He is a fellow of SPIE and has been awarded Honored Science Worker of the Russia, SPIE Educator Award, and FiDiPro (Finland).

**Yong Jeong** is an associate Professor of Department of Bio and Brain Engineering at KAIST and Deputy director of KI for Optical Science and Engineering. Dr. Jeong received his M.D degree in 1991 and Ph.D in neurophysiology in 1997 at Yonsei University. His research fields are Cognitive Neuroscience, Clinical Neurology (degenerative disease, vascular disease), Functional Neuroimaging, and Bioengineering (biosignals). He is also a Neurologist and affiliated with Samsung Medical Center.

**YongKeun Park** is Associate Professor in the Department of Physics, KAIST. Dr. Park received his Ph.D. degree in Medical Physics and Medical Engineering from Harvard-MIT Health Science and Technology in 2010, and joined KAIST. Dr. Park is an Associate Editor of Experimental Biology and Medicine, and Editorial Board Member of Scientific Reports, Optics Express, and Journal of Optical Society of Korea. To learn more about his research projects, visit his website: http://bmol.kaist.ac.kr

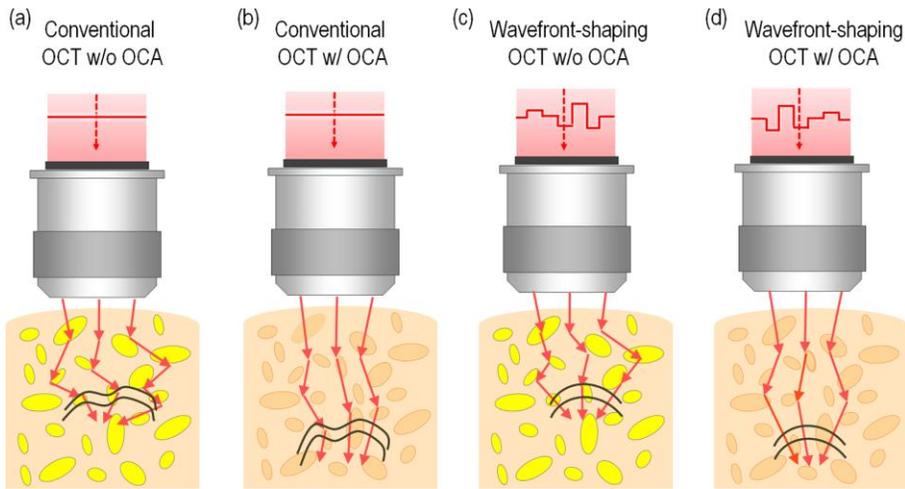

**Fig. 1** Principle of optical clearing agents and wavefront shaping in an OCT system. (a) Conventional OCT imaging. (b) After the application of OCAs, the overall scattering of the tissue is reduced. (c) Wavefront shaping in OCT. The optimal incident wavefront generates a tight focus inside the tissue. (d) Combined effect of OCAs and wavefront shaping. Maximal penetration depth can be achieved with reduced scattering of the sample and optimal incident light field.

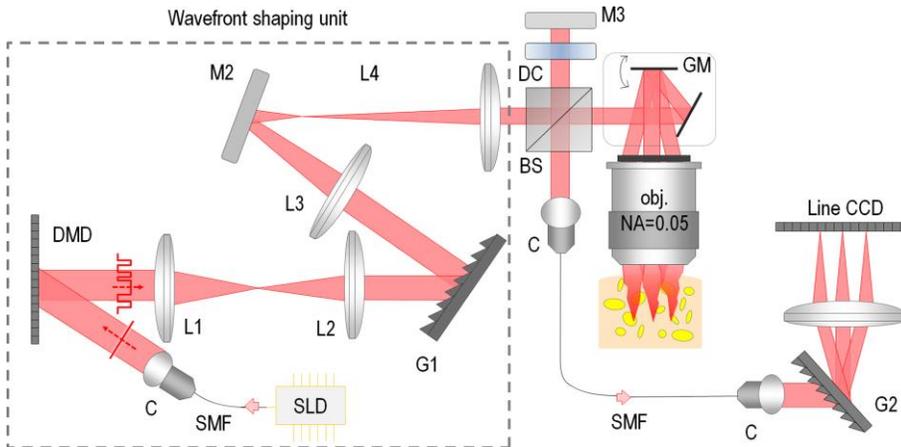

**Fig. 2** Experimental setup. Gray dashed rectangular area: wavefront shaping unit: SLD - super-luminescent diode, DMD - digital micro-mirror device, G1 - dispersion compensating grating; BS - beam splitter, DC - dispersion compensator, GM - galvo mirror, G2 – spectral OCT grating; C – collimators, L - lenses, M - mirrors, SMF – single mode fiber. The intensity modulation after the DMD is not to scale.

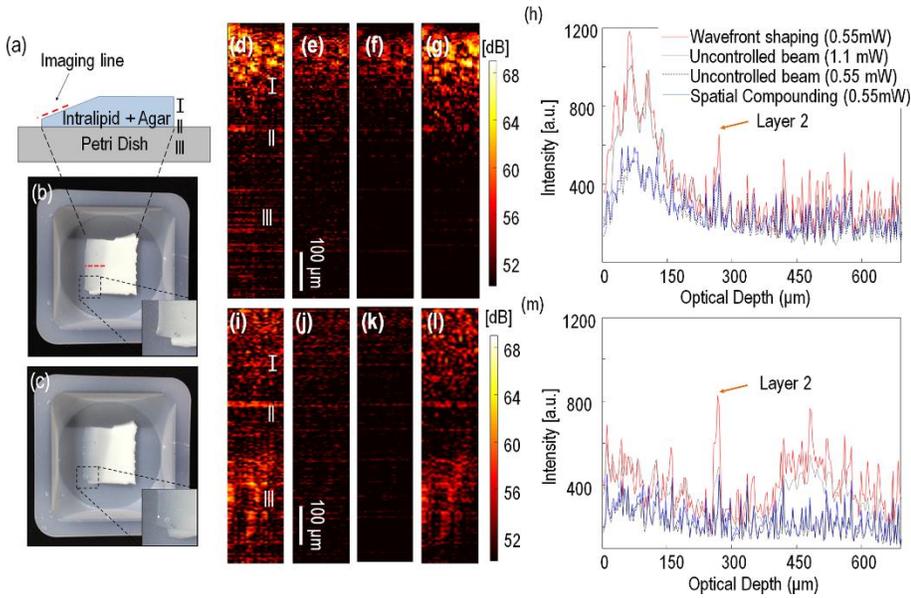

**Fig. 3** (a) Structure of a tissue phantom. Photograph of the tissue phantom (b) before applying the OCA and (c) after applying the OCA. (inset) magnified views of the edges. (d-h) Images acquired before the OCA. (i-m) Images acquired after the OCA. (d,i) Optimized image obtained by wavefront control with input power of 0.55 mW. (e,j) Images acquired for an uncontrolled beam of 0.55 mW. (f,k) Images obtained by spatial compounding method. (g,l) Images acquired for uncontrolled beams of 1.1 mW. (h,m) The averaged A-scan profiles along 25 A-scans were plotted for each method.

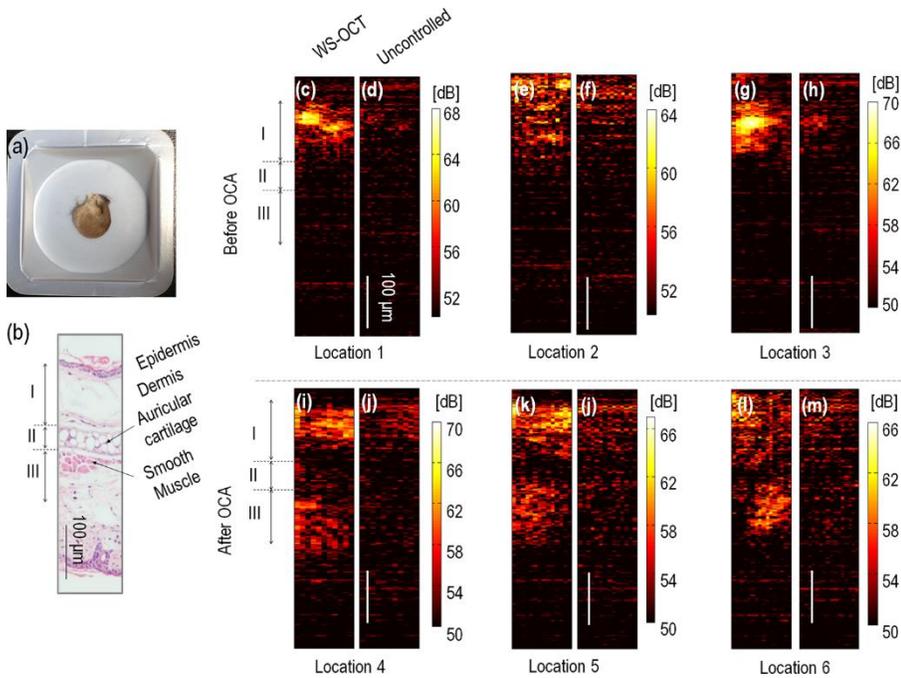

**Fig. 4** (a) Dissected mouse ear; (b) histology of the ear; (c-h not shown) images acquired before the OCA; (i-n not shown) images acquired after the OCA; (c,e,g,i,k,m not shown) image optimized by wavefront control with the input power of 0.55 mW at location 1-6; (d,f,h,j,l,n not shown) images acquired for uncontrolled beam of 0.55 mW at locations 1–6.

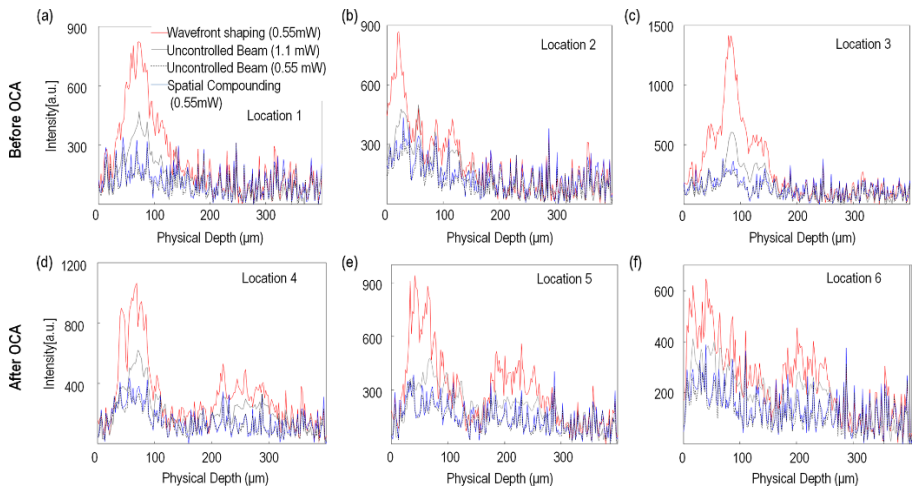

**Fig. 5** The averaged A-scan profiles along 25 A-scans for each method (a-c) before the OCA and (d-f) after the OCA at location 1–6.